\documentstyle[epsfig,aps,floats]{revtex}
\begin{document}
\tighten
%
%
\title{HFB theory for nuclei near the drip-lines: continuum coupling}

%
\author{V.~E.~Oberacker, A.~S.~Umar, J.~Chen, E.~Teran}
\address{
Department of Physics \& Astronomy, Vanderbilt University,
   Nashville, TN 37235, USA}

%
%
\maketitle

\begin{abstract}

We have developed a new HFB code that
specifically addresses nuclear structure physics near the driplines.
The HFB equations are solved on a two-dimensional lattice for axially
symmetric even-even nuclei using B-Spline techniques. The quasiparticle
energy spectrum is obtained by direct diagonalization of the 
HFB lattice Hamiltonian with LAPACK. The energy spectrum extends 
high into the continuum, up to several thousand MeV. Calculations with
Skyrme forces and (density-dependent) delta pairing interactions
are now underway.

\end{abstract}

%
%
\section{Introduction}

Near the neutron or proton drip lines, large pairing correlations are
expected which can no longer be described by a small residual interaction. 
Furthermore, the outermost nucleons are weakly bound (which implies a
large spatial extent), and they are strongly coupled to the particle
continuum. These features represent major challenges for the mean
field theories. We solve the self-consistent mean field plus pairing
(Hartree-Fock-Bogoliubov) equations for axially symmetric
even-even nuclei on a two-dimensional lattice. We utilize the
Skyrme M$^*$ interaction, and pairing is described either by a pure delta
force or by a density-dependent delta interaction (DDDI).

\section{Numerical method and preliminary results}

High numerical accuracy is achieved by representing the operators and
wavefunctions in terms of Basis-Splines; a combination of the Galerkin
and collocation method is utilized \cite{OU99}. Our work represents a natural
extension of the 1-D calculations for spherical nuclei by Dobaczewski
et al. \cite{DN96}. 

\begin{figure}
\begin{center}
\includegraphics[scale=0.33]{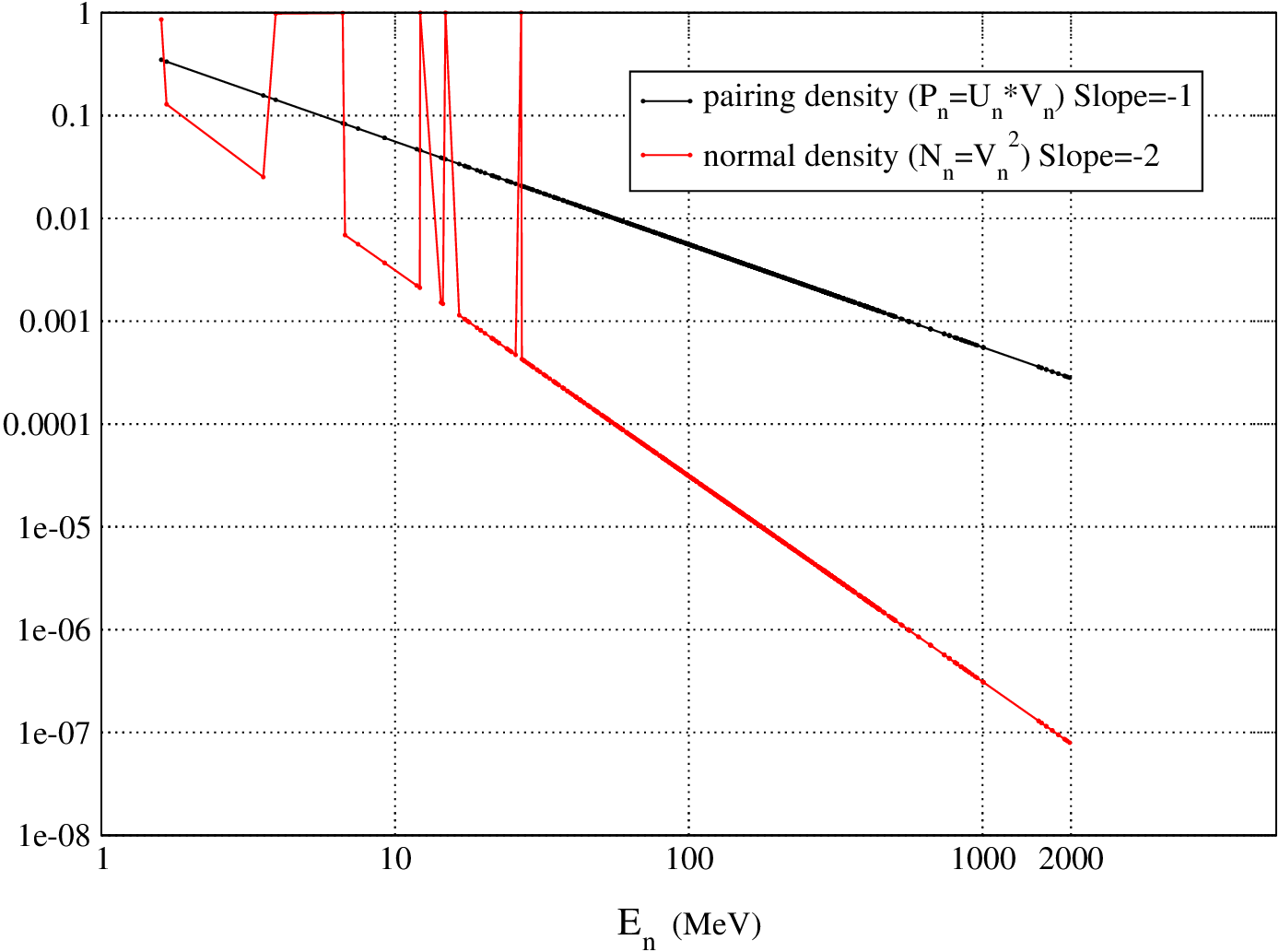}
\end{center}
\caption{Test of HFB code for constant pairing Hamiltonian: all
observables are identical to HF + BCS, as expected.}
\label{ne22_en_bcs}
\end{figure}

We have successfully tested our numerical algorithm by solving the HFB equations
for a constant pairing Hamiltonian in which case the
problem becomes equivalent to HF + BCS. As an example,
we show in Fig.~\ref{ne22_en_bcs} the quasiparticle energy spectrum of the
normal and pairing density for $^{22}$Ne. The spectrum of the lattice HFB Hamiltonian
is obtained by direct diagonalization 
with LAPACK. In this way, we obtain the whole quasiparticle energy
spectrum at once, up to about $E_n=2000$ MeV. In calculating observables, we
cut off this spectrum at an equivalent s.p. energy of about 60 MeV. A correct
representation of high-energy continuum states is crucial for physics near the driplines.

We wish to emphasize that this numerical test is non-trivial because the HFB code
solves a \emph{quasiparticle} energy spectrum (upper and lower continuum, no lower
bound) while the HF + BCS codes solve for the energy spectrum of \emph{real particles} 
(one upper continuum only, lower bound). 

In the following, we discuss preliminary results obtained with the full HFB code.
All calculations are for $^{22}$Ne with the Skyrme M$^*$ interaction in the p-h
channel and a pure delta pairing interaction (strength $V_0=-173 MeV fm^3$, taken
from ref. \cite{DN95}). We utilized B-Spines of order $M=7$, a lattice size of
8 fm in radial direction, and a lattice spacing of about 1 fm.

\vspace{1.0cm}

\begin{figure}[h]
\begin{center}
\includegraphics[scale=0.33]{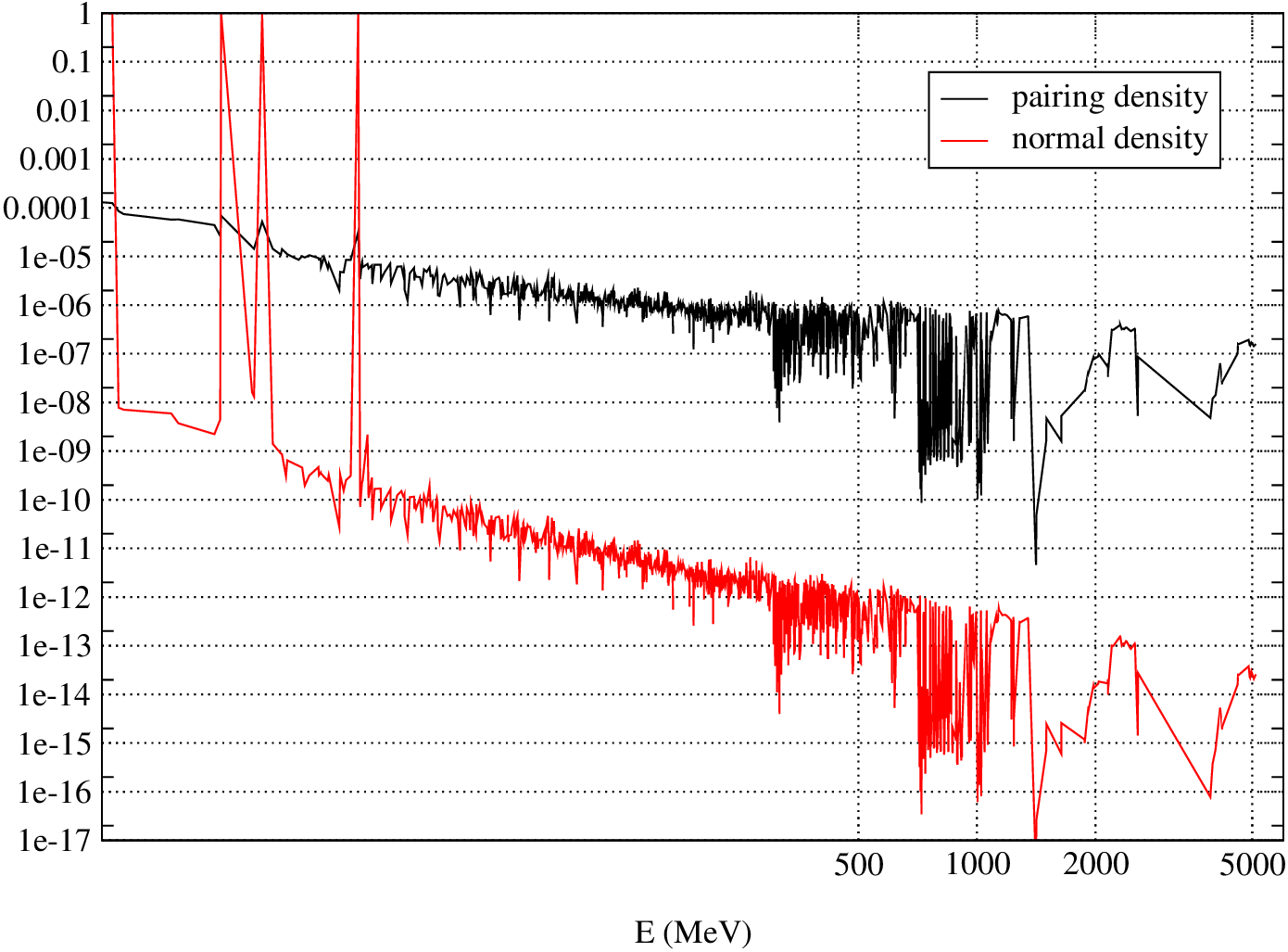}
\end{center}
\caption{HFB code with SkM* and pairing delta interaction: quasiparticle
energy spectrum of densities}
\label{ne22_en_hfb}
\end{figure}
Fig.~\ref{ne22_en_hfb} shows that both the normal density and the pairing
density develop more structure in the quasiparticle energy spectrum, as
compared to the BCS pairing case. This is in agreement with the results
from 1-D calculations.

\vspace{1.0cm}

\begin{figure}[h]
\begin{center}
\includegraphics[scale=0.3]{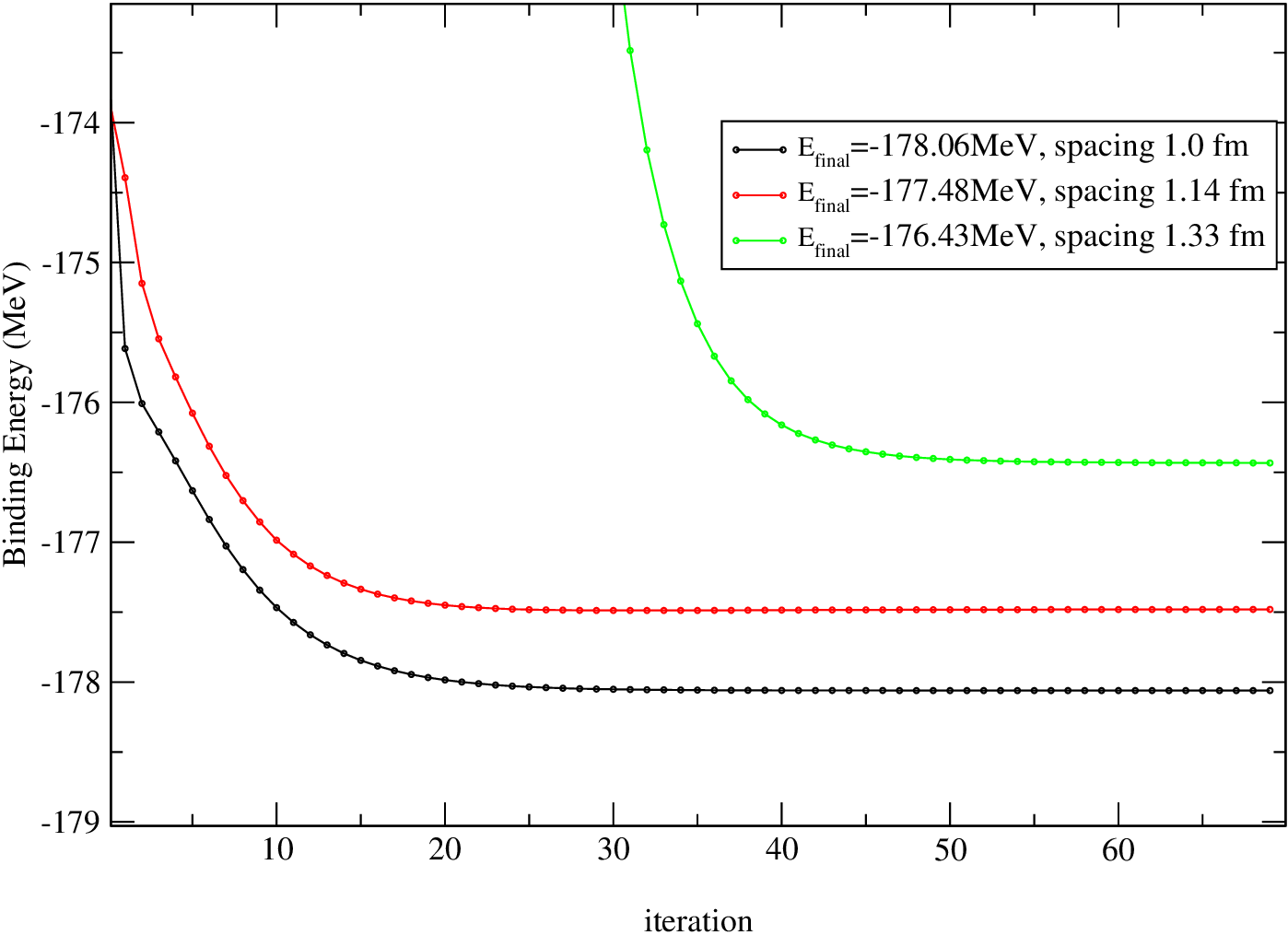}
\end{center}
\caption{Convergence of total nuclear binding energy
as function of number of HFB iterations in LAPACK}
\label{hfb_etot}
\end{figure}
 Fig.~\ref{hfb_etot} demonstrates the convergence of the total
nuclear binding energy as a function of the number of iterations. The three
different curves correspond to lattice spacings of $\Delta x = 1.33, 1.14, 1.00$ fm.
\begin{figure}[t]
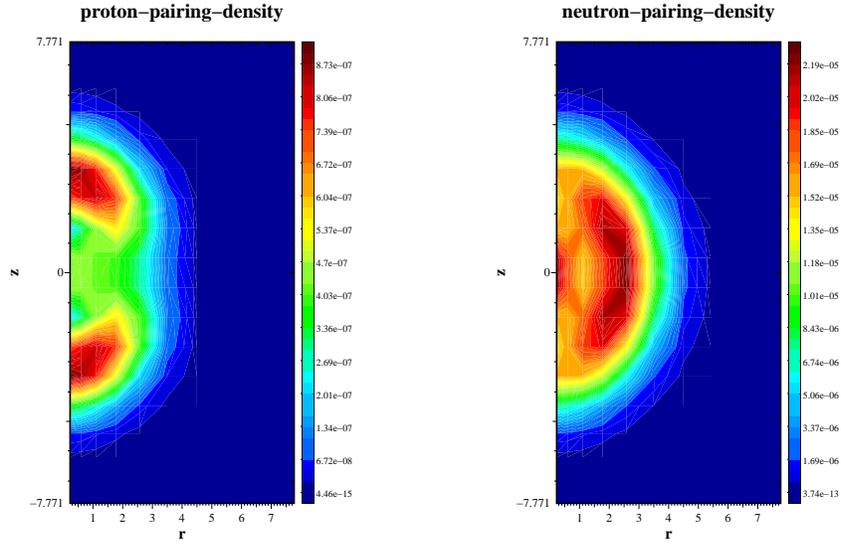

\begin{center}
\includegraphics[scale=0.3]{ne22_pairdens_prot.epsi}
\includegraphics[scale=0.3]{ne22_pairdens_neutr.epsi}
\end{center}
\caption{proton and neutron pairing density for $^{22}$Ne}
\label{ne22_pairdens}
\end{figure}
Fig.~\ref{ne22_pairdens} shows contour plots of the pairing density for
protons and neutrons. The square of the pairing density describes the probability
of finding a \emph{correlated} nucleon pair with opposite spin directions at position 
${\bf r}$ \cite{DN96}.

We are now in the process of adjusting the DI and DDDI pairing force parameters
to the tin isotope chain and plan to do a detailed comparison for spherical
nuclei with Dobaczewski's 1-D HFB code \cite{DN96}. After
these tests are finished, we will be in a position to address the
new physics near the drip lines such as neutron halos and neutron skins,
proton radioactivity etc. We are going to calculate binding energies, neutron and proton
separation energies, pairing gaps, normal and pairing densities, rms radii, and
electric or magnetic moments.
 
We have adopted the Skyrme force parameterization
suggested in ref.\cite{RD99}. Currently, all of our HFB calculations are
carried out with the SkM$^*$ interaction. In the near future, we plan to
investigate some of the more recent parameterizations, in particular
SLy4 and to explore the sensitivity of observables to these forces.
We will also add constraints ($Q_{20}, Q_{30}, \omega j_x$)
to the HFB code in order to calculate potential energy surfaces for
nuclei far from stability. Our long-range plan is to extend the current ground state
theory to excited states, i.e. quasiparticle RPA on the lattice \cite{Wi00}.

%
%
\acknowledgements
This work was supported by the U. S. Department of Energy 
under contract DE-FG02-96ER40975 with Vanderbilt University.
Some of the numerical calculations were carried out on CRAY-SV1
supercomputers at NERSC.
%
%
\bibliographystyle{try}

%
%
\end{document}